\documentclass[12pt]{article} % single column, double spaced

\usepackage{ol2}

\begin{document}

%\twocolumn[ %% activate for two-column option

\title{Physical picture, pattern-control, and detection approach for
tightly focused beams: In the view of Fourier optics}
%
%% For REVTeX it is possible to automate superscript and e-mail callouts with the superscriptaddress option; see REVTeX4 documentation.

\author{Daquan Lu, and Wei Hu}
\address{
Laboratory of Photonic Information Technology, South
China Normal University, Guangzhou 510631, China\\
}

\begin{abstract}
We investigate the mechanism of the nonparaxial propagation of the
tightly focused beams in the view of Fourier optics. It shows that
it is the phase of the angular spectrum which induces the
interesting evolution of the tightly focused beams. Based on the
theory of Fourier optics, we propose an approach on controlling and
detecting the focusing spot of the tightly focused beams.
\end{abstract}
\maketitle %% null function with osajnl.sty

It is well-known that there are two types of propagation theory for
laser beams, i.e. the paraxial propagation theory and the
nonparaxial propagation theory. The paraxial theory is able to give
an accurate description while the divergence angles of beam is very
small and the beam width is much larger than its wavelength.
However, if a beam is with a large divergence angle or is tightly
focused, the paraxial approximation is invalid and it requires a
non-paraxial correction to the paraxial solution.

In the past decades, the paraxial diffraction equation has been
thoroughly investigated. Various types of beam solutions with
different transverse profiles have been obtained in Cartesian,
circular cylindrical, and elliptical coordinates (e.g., \cite{1,2,3}
and references therein). These solutions can be roughly classified
into two types: i) shape-invariant beams, such as LG, HG, and IG
beams; and ii) shape-variant beams, such as higher-order
elegant-Hermite-Gaussian (EHG), elegant-Laguerre-Gaussian (ELG), and
elegant-Ince- Gaussian (EIG) beams. The propagation of these beams
has been investigated in detail and many parameters, such as width,
divergence, radius of curvature, and quality factor, have been
introduced to describe their propagation. In summary, the theory of
paraxial propagation has been well developed over the past decades.
On the other hand, there are also various approaches for the
nonparaxial propagation of beams\cite{4,5,6,7,8,9}.
 However, because of the complexity of the nonparaxial wave equation,
the mechanism for the beam evolution under the nonparaxial condition
is  still unclarified.

In this paper, we investigate the mechanism of the nonparaxial
propagation in the view of Fourier optics. It shows that it is the
phase of the angular spectrum which induces the interesting
evolution of the tightly focused beams. Based on the theory of
Fourier optics, we propose an approach on controlling and detecting
the focusing spot of the tightly focused beams.

The studies of this paper is based on the scalar wave equation,
\begin{equation}
(\nabla_\bot^2+\partial_z^2-2ik\partial_z)E=0,
\end{equation}
which describes the evolution of the electromagnetic field in free
space. However, it is well known that, outside the realm of paraxial
approximation, the electromagnetic field have to be handled as a
vectorial one since the coupling between each components. And the
longitudinal component is not negligible for a nonparaxial pulsed
beam. Fortunately, in free space, the longitudinal component can be
evaluated from the knowledge of the transverse part, i.e., base on
the relation $\nabla\cdot E=0$. In this perspective and also for
conciseness,  we only consider the transverse components of the
electromagnetic field and our results are limited to the scalar
field in this paper.

A scalar laser field $E(x,y,z)$ can be regarded as the superposition
of various planar waves which propagate in different directions.
Every plane wave experiences a corresponding phase shift when a
plane wave propagate from $z=0$ to $z=z$:
\begin{equation}
\tilde{E}(z)=\tilde{E}(0)\exp(i\mathbf{k}\cdot\mathbf{r})
=\tilde{E}(0)\exp(i\mathbf{k_\bot}\cdot\mathbf{r_\bot}+i{k_z}z),
\end{equation}
where $k_z=\sqrt{k^2-k^2_\bot}$,
$\mathbf{k_\bot}=k_x\hat{e}_x+k_y\hat{e}_y$, $k^2_\bot=k^2_x+k^2_y$.
At the plane $z=z$, the linear superposition of all spectrums yields
the field in spatial domain, i.e.
\begin{equation}
{E}(z)=\int\int\tilde{E}(0)\exp(i\mathbf{k_\bot}\cdot\mathbf{r_\bot}
+iz{\sqrt{k^2-k^2_\bot}})
 dk_xdk_y.\label{int1}
\end{equation}
Equation (\ref{int1}) is tantamount to
\begin{equation}
{E}(z)=\hat{F}^{-1}\left\{\hat{F}\left\{{E}(0)\right\}
\exp\left(iz{\sqrt{k^2-k^2_\bot}}\right)\right\}, \label{NP}
\end{equation}
 where operators $\hat{F}$ and $\hat{F}^{-1}$ represent the Fourier
transform and the inverse Fourier transform, respectively.
Therefore, in the view of Fourier optics, the nonparaxial
propagation from $z=0$ to $z=z$ results from three steps: i)
transform the field at $z=0$, i.e., ${E}(0)$, to the angular
spectrum domain; ii) add a phase $\Delta \phi =k_zz$ on
$\tilde{E}(0)$; iii) inversely transform the angular spectrum  at
$z=z$ to the spatial domain.

During propagation, the second step plays an important role since
different angular-spectral components experience different phase
shift $\Delta \phi$. In fact, we can borrow the ideal of the
dispersion of a pulse in fiber to analyze the influence of the
difference of the phase shift on the evolution of the beam. We make
a Taylor expansion on $\Delta \phi$ and have
\begin{equation}
\Delta \phi=\left(\beta_0+\frac{1}{2!}\beta_2
k_\bot^2+\frac{1}{4!}\beta_4 k_\bot^4+...\right)z,\label{dispersion}
\end{equation}
where
\begin{equation}
\beta_n=-\frac{{n!|n-3|!!}}{{n!!k^{n-1}}}.
\end{equation}
As shown in Eq. (\ref{dispersion}), the beam experiences the spatial
dispersion during propagation. It is noted that there are two
critical difference between the temporal dispersion of the pules in
fiber and the spatial dispersion of the beam in free space: i) The
beam experiences only even-order spatial dispersion, whereas the
pulse experiences both even- and odd-order temporal dispersion.
ii)For different media of the fiber and wavelengths of the pulse,
the temporal dispersion in fiber can be positive or negative, and
the dispersion with different orders can be with different signal.
However for the diffraction of the beam in free space, the signal of
all orders of spatial dispersion ($\beta_n<0$) are negative, which
means the higher spatial frequency experiences lager phase-shift
than the lower spatial frequency. During propagation, all orders of
spatial dispersion together cause a negative spatial chirp (or in
other words, a convex copahsal surface).

 If the beam width is much larger than the wave lenth
 and the divergence angle is very small (or in other words, the paraxial approximation is
 satisfied), the
 angular-spectral width so narrow that only the lowest dispersion
$\beta_2$ is necessary to be taken into account. under this
condition  Eq. (\ref{NP}) reduces to
 \begin{equation}
{E}^{(p)}(z)=\int\int\tilde{E}(0)\exp\left(i\mathbf{k_\bot}\cdot\mathbf{r_\bot}+i\beta_0
z+\frac{i} {2}\beta_2z\right)dk_xdk_y.\label{fresnel}
 \end{equation}
 Eqation (\ref{fresnel}) is called the Fresnel integral, which
 governs the evolution of an arbitrary paraxial beam.

Whereas, if the beam is so narrow that the beam width is comparable
with the wavelength or the divergency angle is very large so that
the angular-spectral  width $\Delta k$ is no longer small enough to
justify the truncation of the expansion (\ref{dispersion}) after the
$\beta_2$ term, the higher orders  of dispersion  should be included
in the expansion. And Eq. (\ref{int1}) deduces to
\begin{equation}
{E}^{(np)}(z)=\int\int\tilde{E}(0)\exp\left(i\mathbf{k_\bot}\cdot\mathbf{r_\bot}
+i\beta_0
z+\frac{i} {2}\beta_2z+\frac{i}
{24}\beta_4z+...\right)dk_xdk_y.\label{npint}
 \end{equation}
Eq. (\ref{npint}) describes the propagation of the nonparaxial
beams, such as the largely divergent beams from the sources whose
size is about the wavelength and the tightly focused beams.

In the following, we will discuss the influence of different orders
of spatial dispersion on the propagation of a tighyly focused beam.
In application, the paraboloids or the spherical lenses provide
positive linear spatial chirp and are frequently used to focus a
beam; and the field next to the lens is with a concave spherical
cophasal surface. In the framework of the paraxial theory, there are
two key characters of the propagation. i) The evolution of the beam
is induced only by the 2nd order spatial dispersion, which does  not
vary the angular spectrum distribution,
 but  induces a spherical phase distribution in the angular spectrum
domain during propagation. In the view of Fourier Optics, the
Hermite-, laguerre-, and Ince-Gaussian functions are the eigen
functions with spherical cophasal surface of the Fourier transform,
i.e.
\begin{equation}
\hat F \{{\Lambda(k_\bot/a_1)}\exp(-ib_1k_\bot^2)\}=
{\Lambda(r_\bot/a_2)}\exp(-ib_2r_\bot^2),
 \end{equation}
 where $\Lambda (\cdot)$ represents an arbitrary Hermite-, laguerre-, or
  Ince-Gaussian function. Therefore, an arbitrary beam, which is resulted from
  the linear superposition of these three types of beams with the same
  waist location and Rayleigh distance, would remain shape-invariant during
  propagation under the paraxial approximation.
The paraxial propagation varies only the beam width and the radius
of the cophasal surface. ii) Because only the 2nd order spatial
dispersion is taken into account and the higher orders of dispersion
are  neglected, the dispersion-induced negative linear chirp can
completely cancel the initial positive linear chirp at a certain
plane. According to Eq. (\ref{int1}), at that plane the beam width
is Fourier-Transform-limited and arrives its minimum, thus that
plane becomes the waist of the focused beam; and the intensity
distribution is symmetric about the waist (we will call it the
pseudo-waist in the following, because it is not really the waist
for the tightly focused case).

  \begin{figure}[t]
\begin{center}
   \leavevmode
   \includegraphics[height=0.5in]{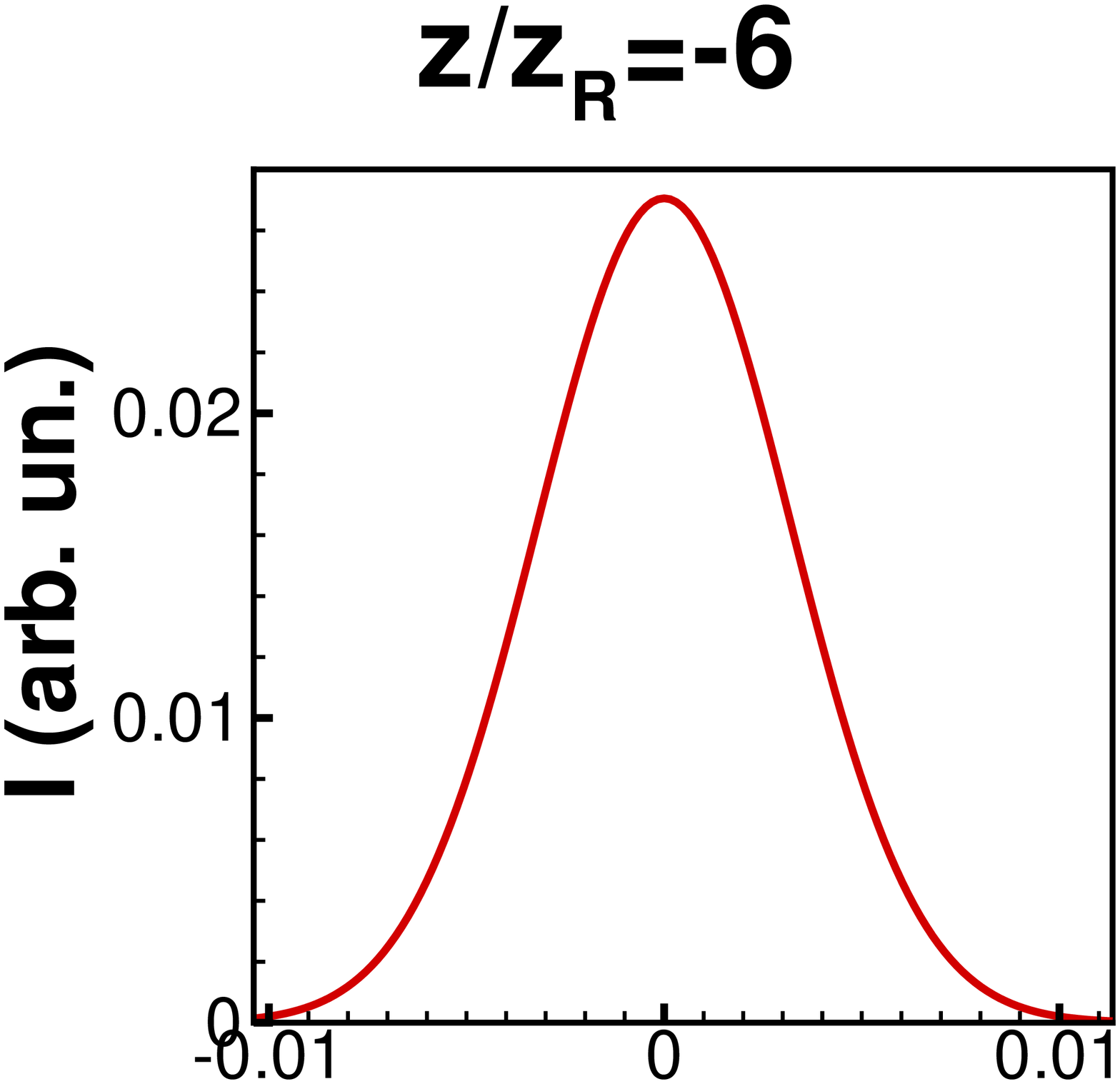}
   \includegraphics[height=0.5in]{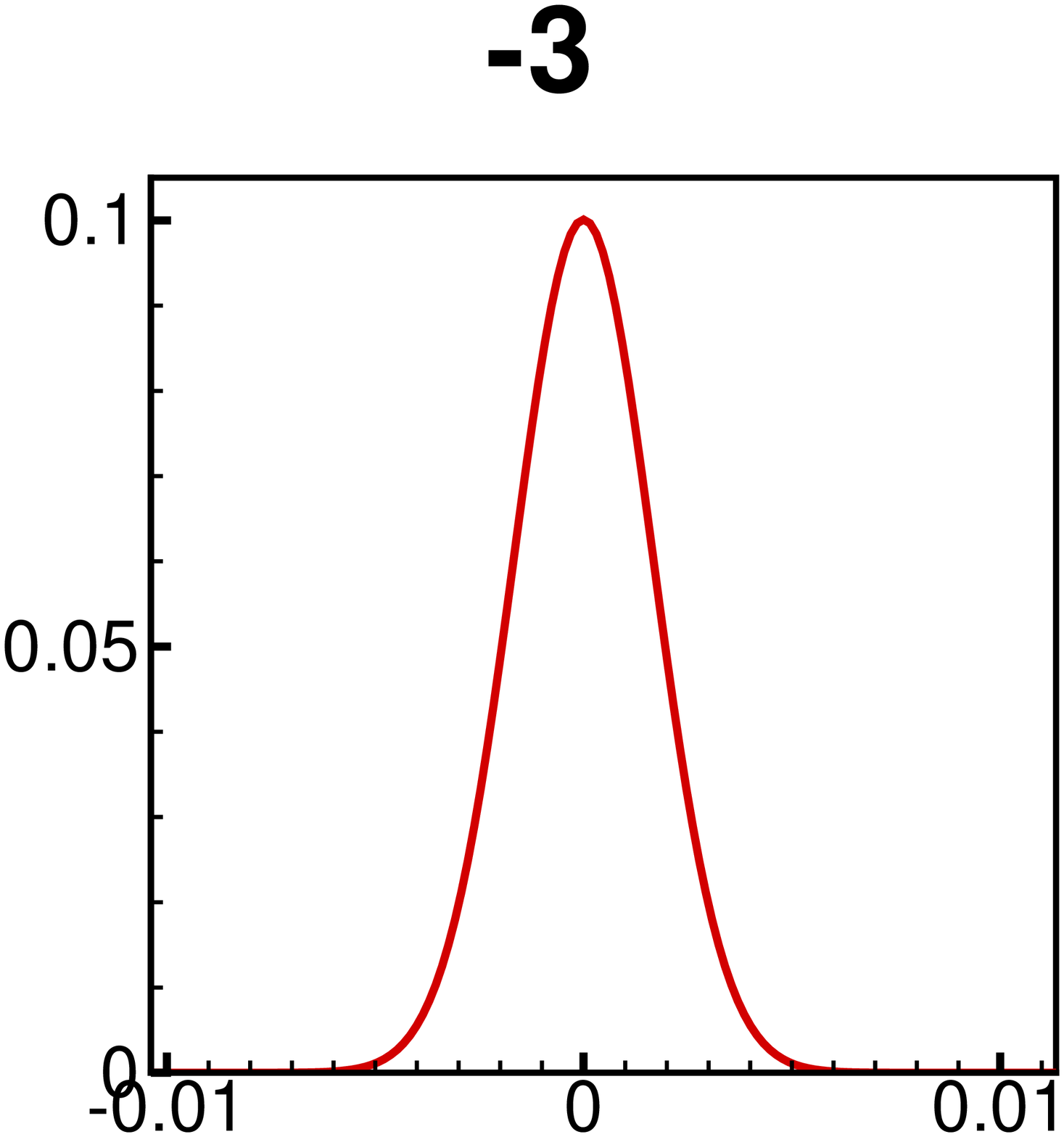}
   \includegraphics[height=0.5in]{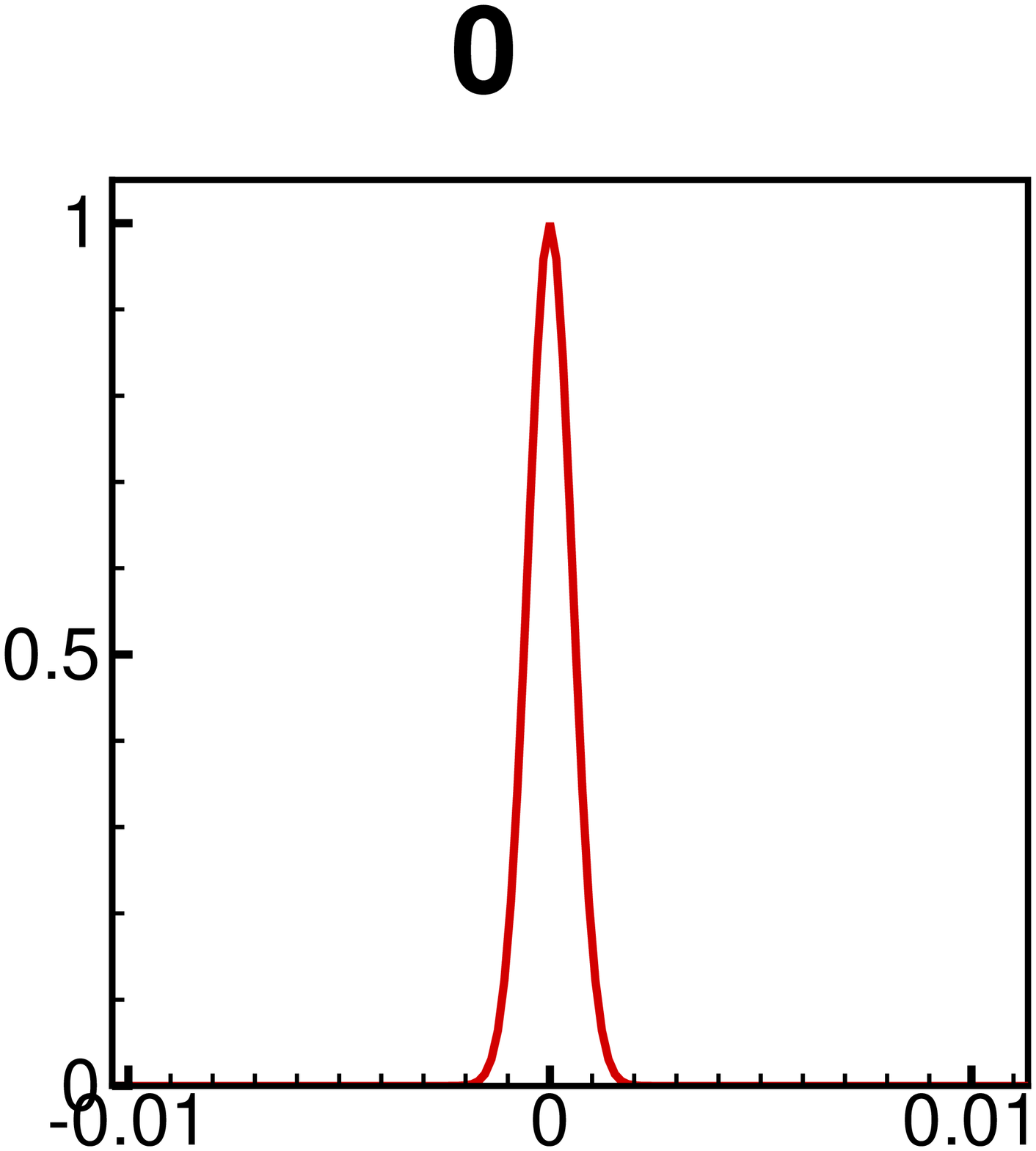}
   \includegraphics[height=0.5in]{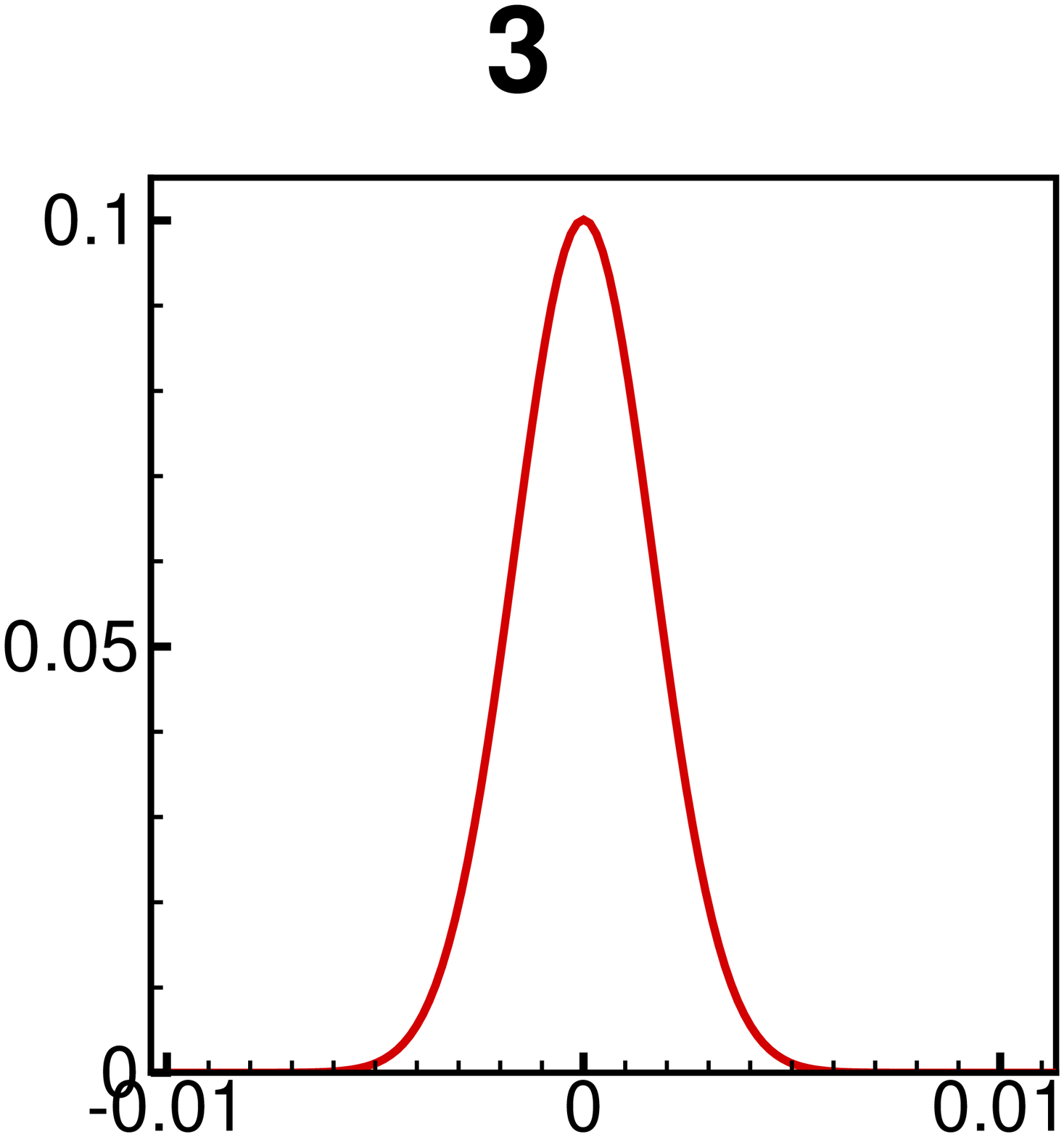}
   \includegraphics[height=0.5in]{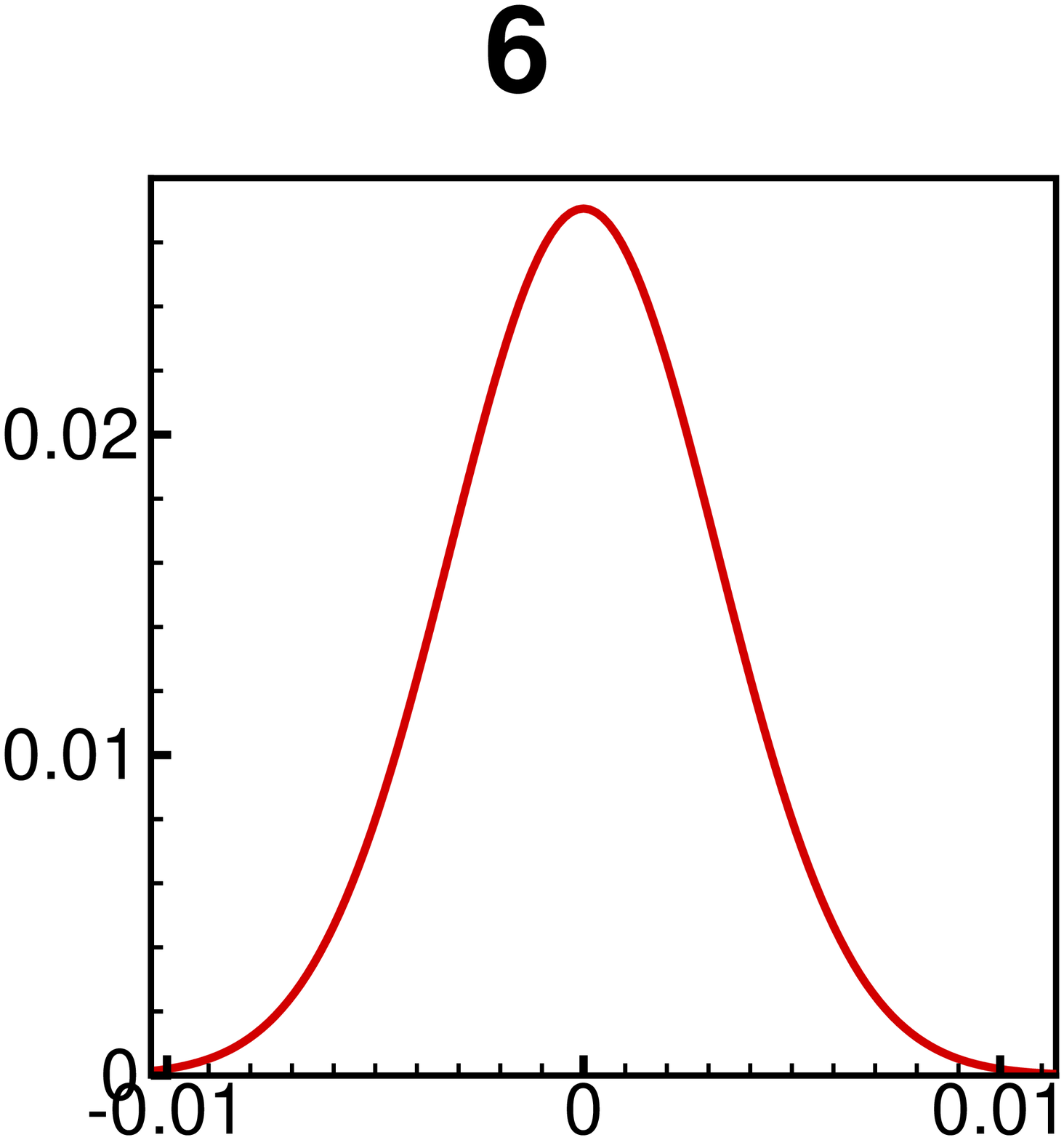}\\
   \includegraphics[height=0.45in]{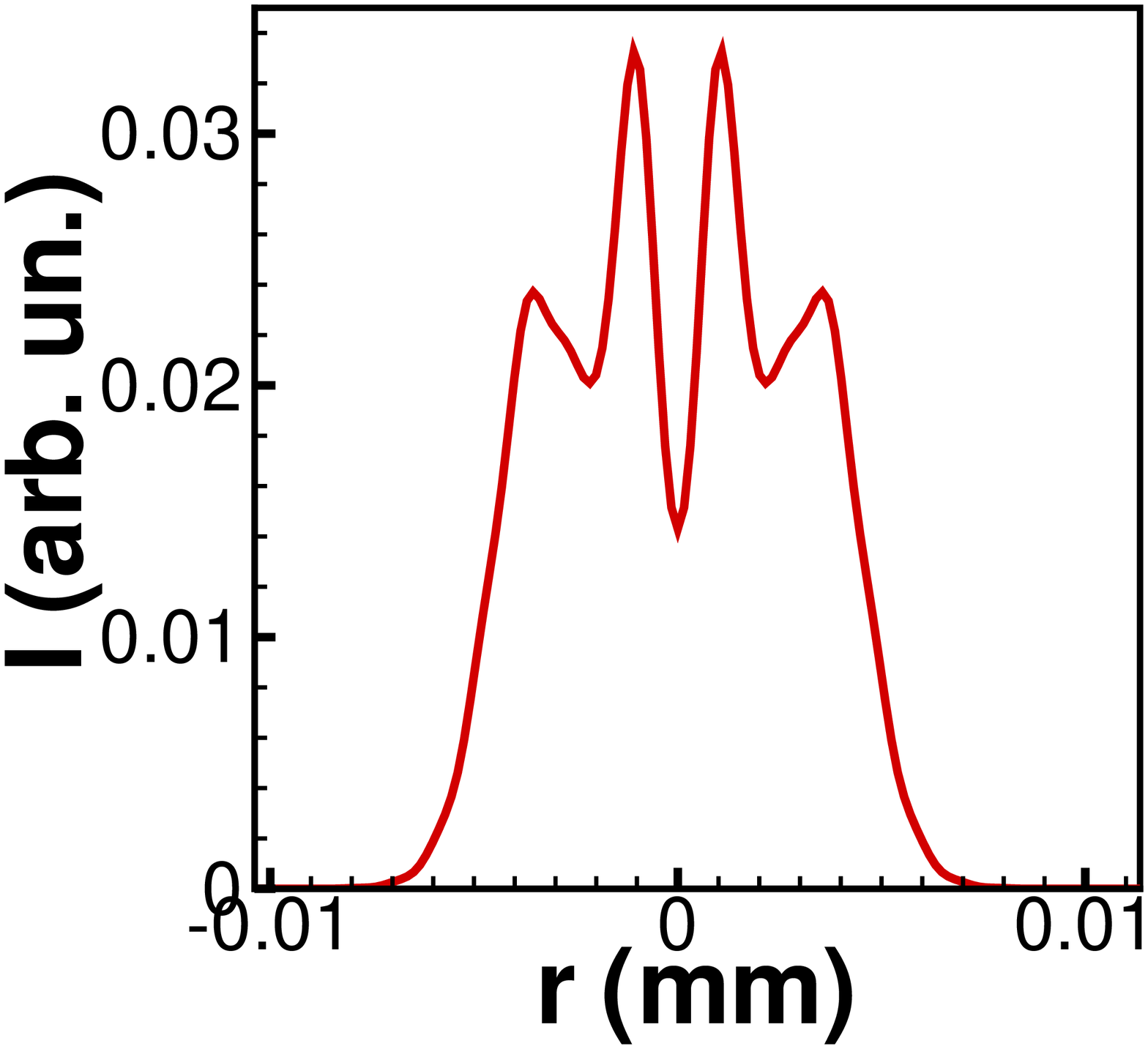}
   \includegraphics[height=0.45in]{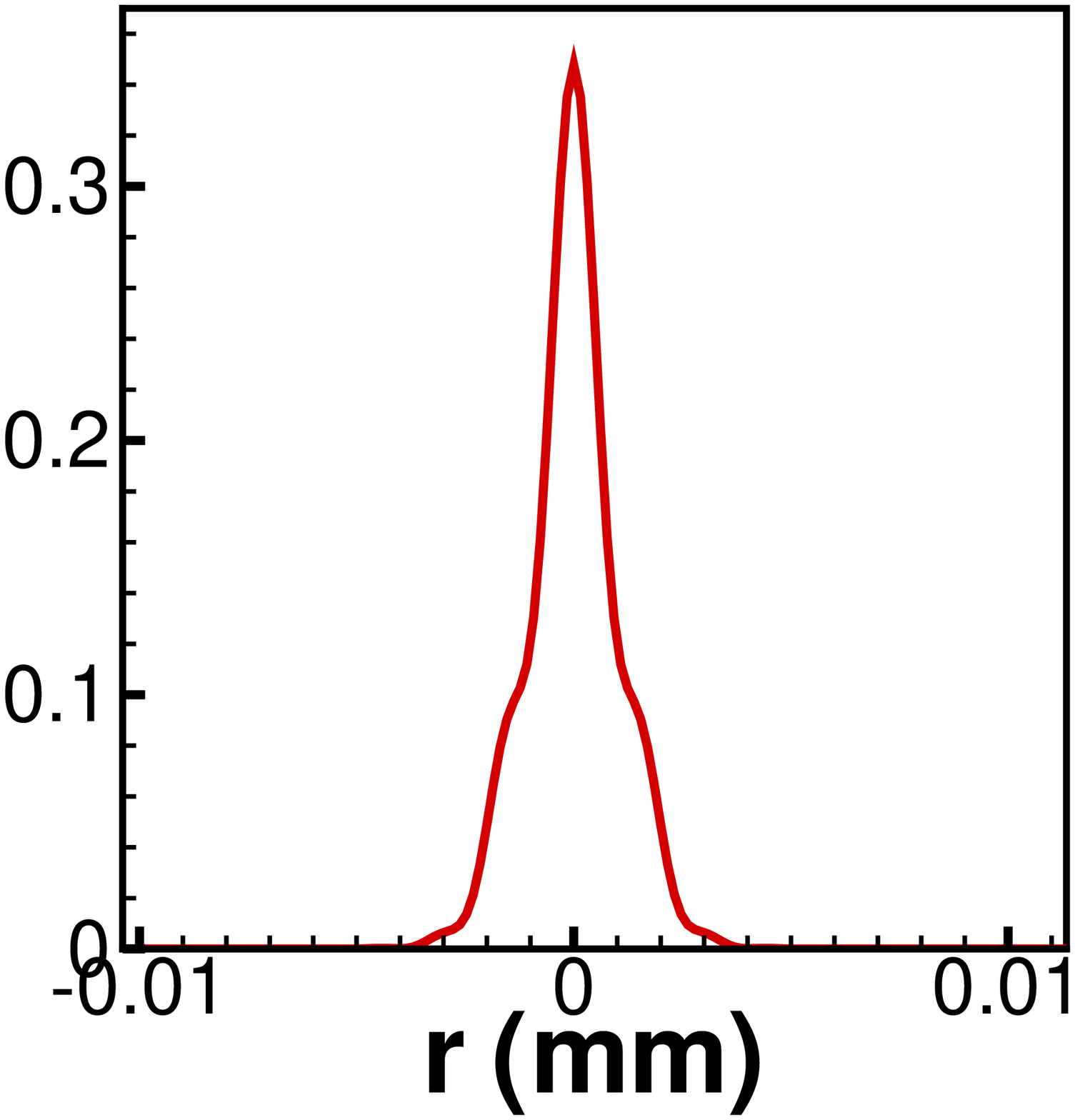}
   \includegraphics[height=0.45in]{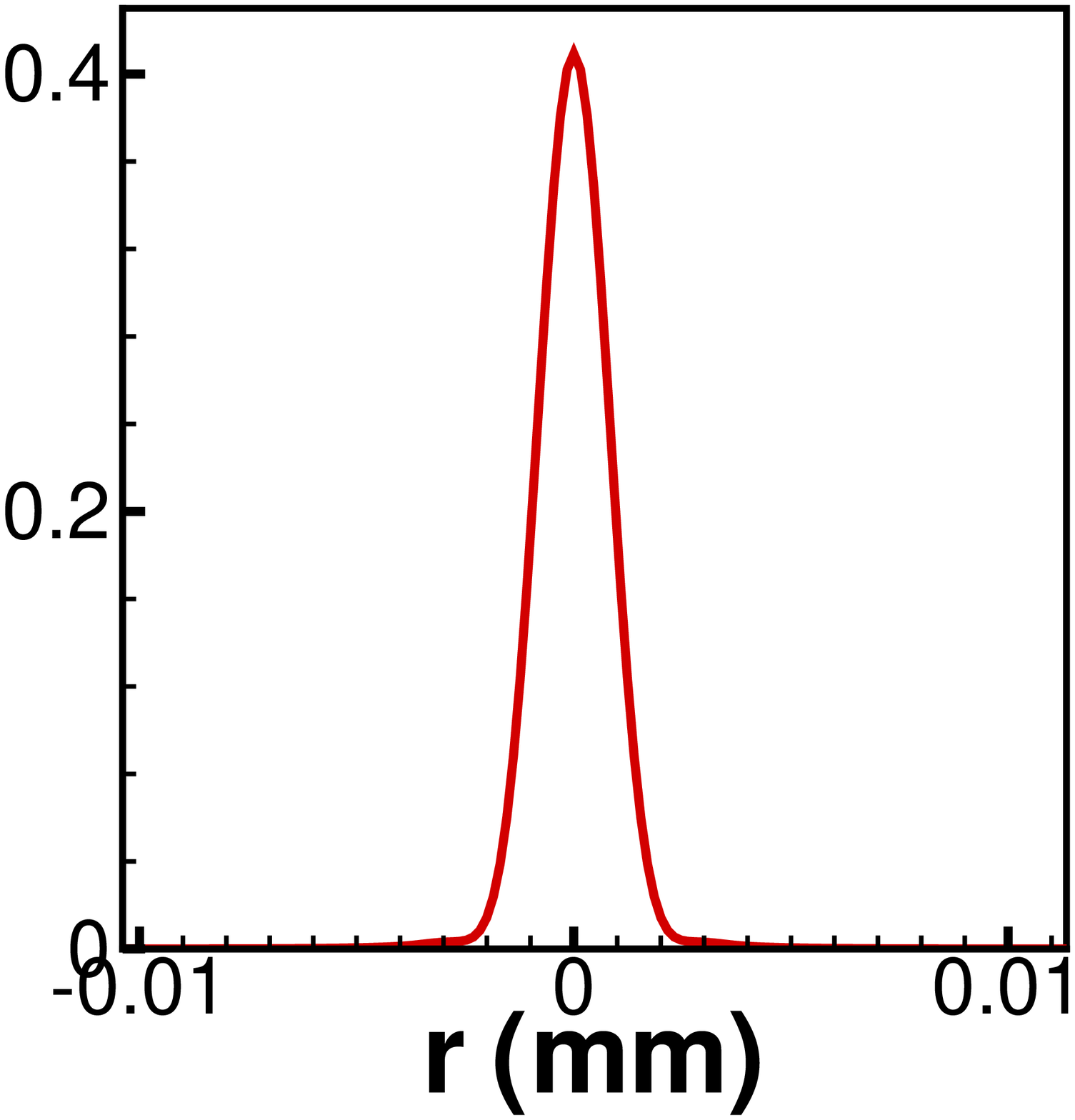}
   \includegraphics[height=0.45in]{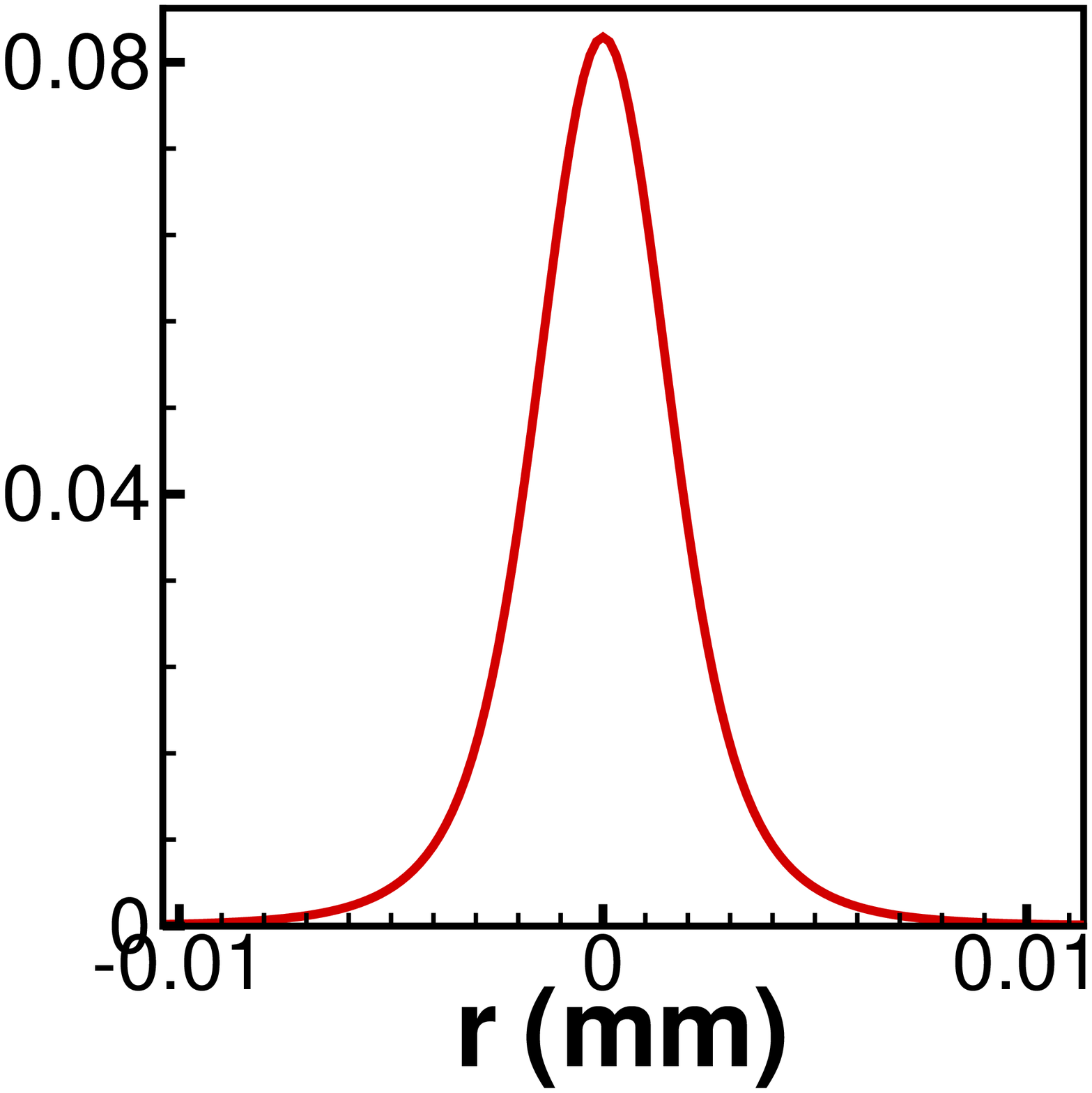}
   \includegraphics[height=0.45in]{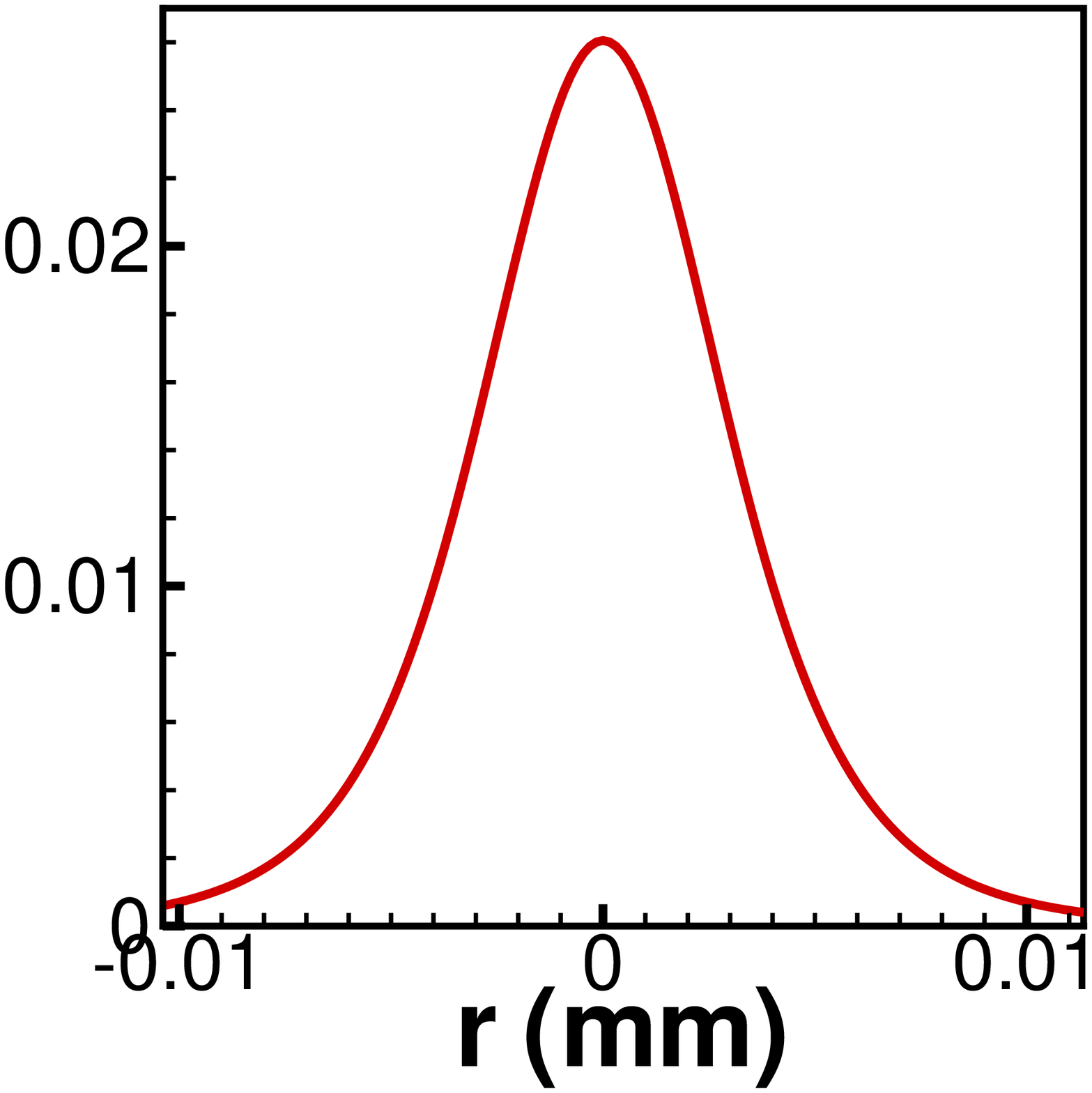}\\
    \includegraphics[height=0.42in]{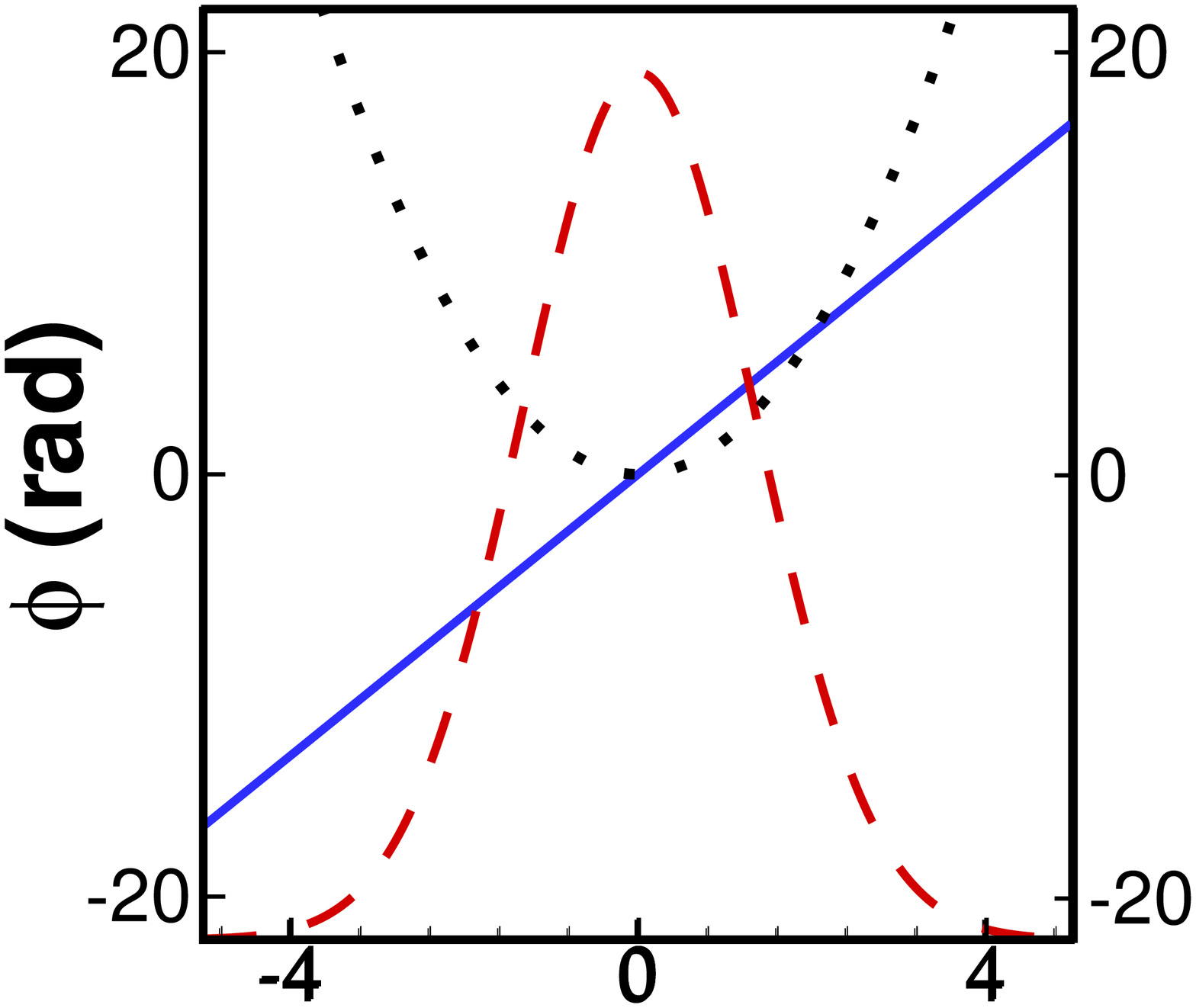}
   \includegraphics[height=0.42in]{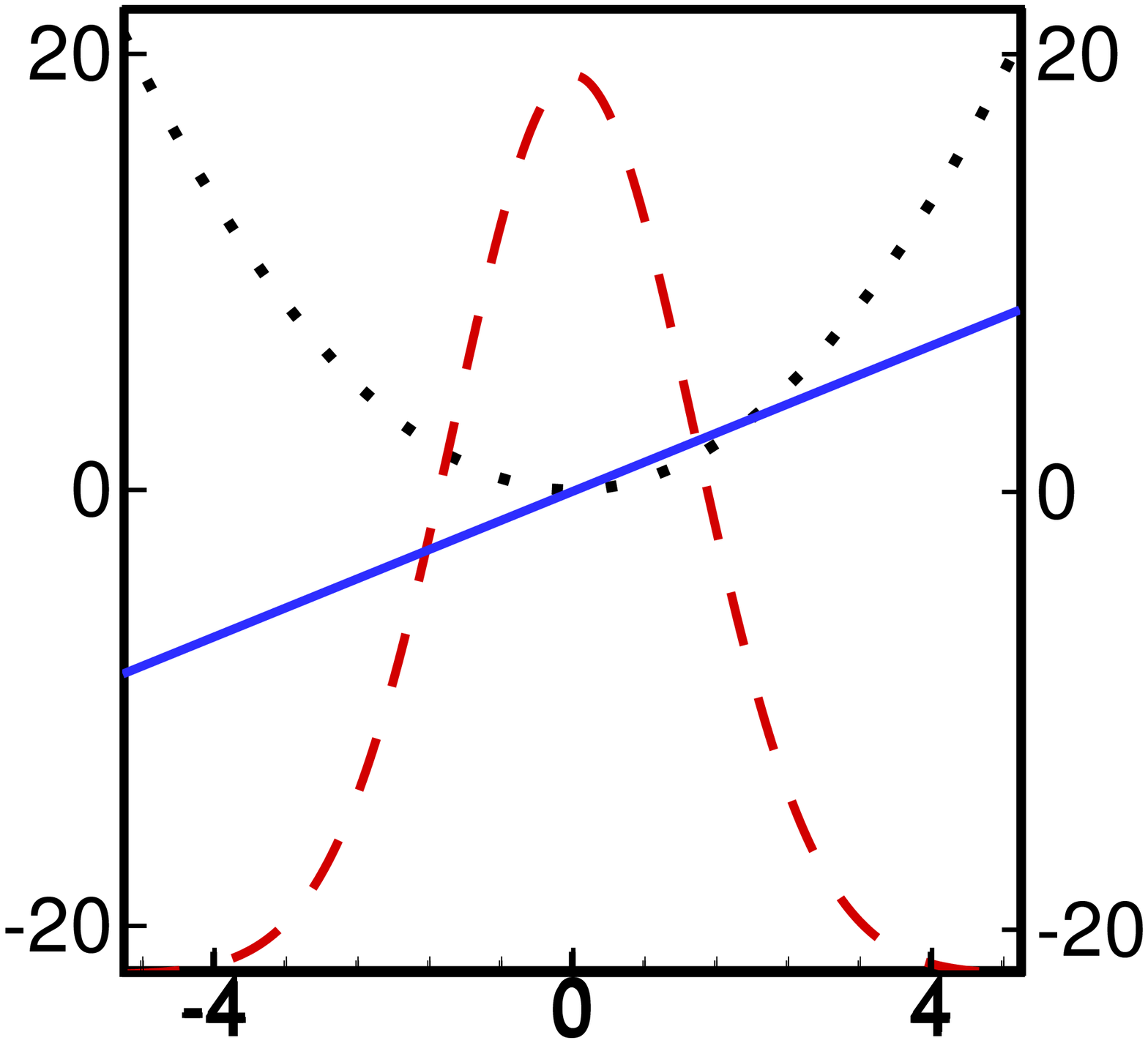}
   \includegraphics[height=0.42in]{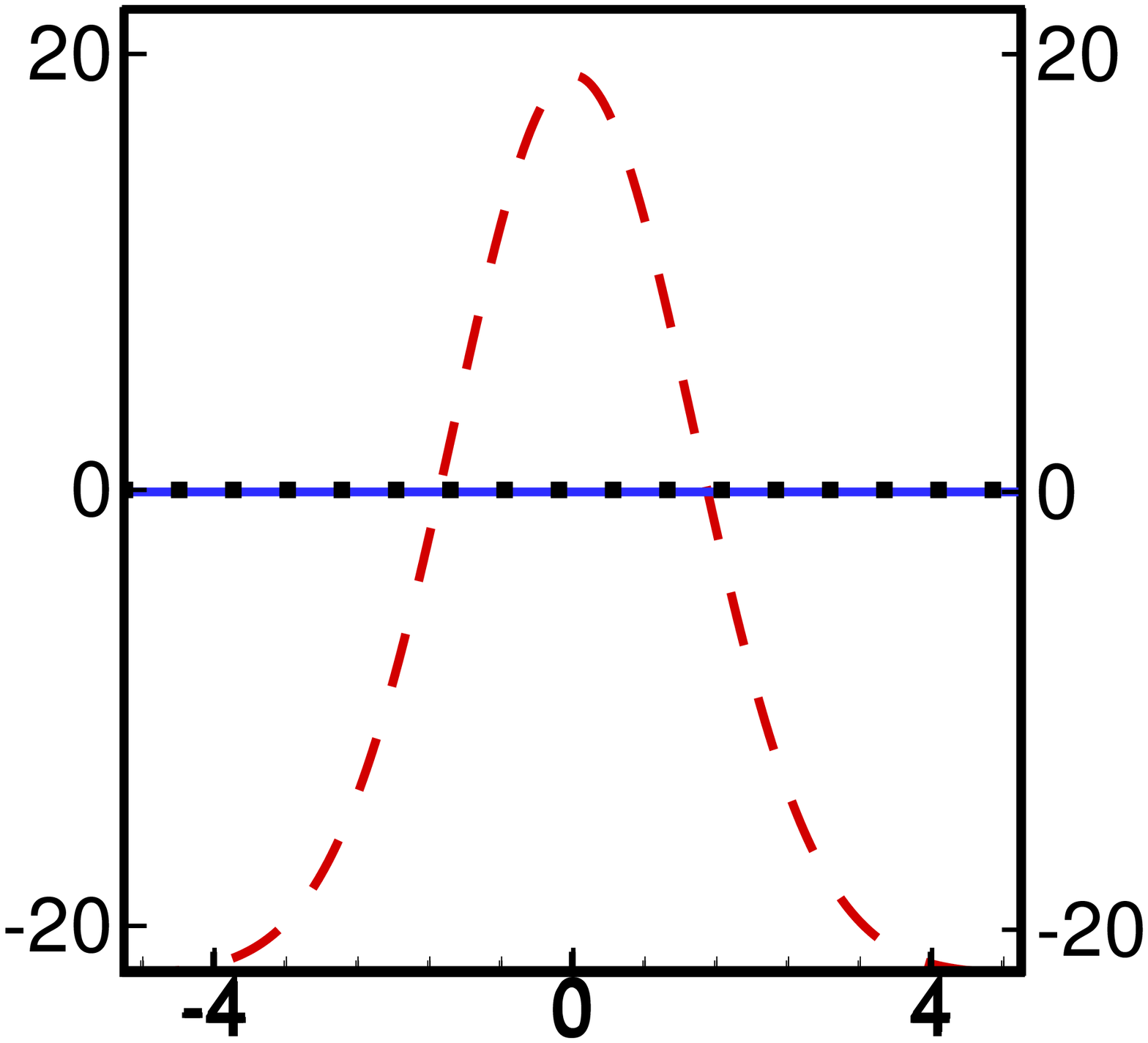}
   \includegraphics[height=0.42in]{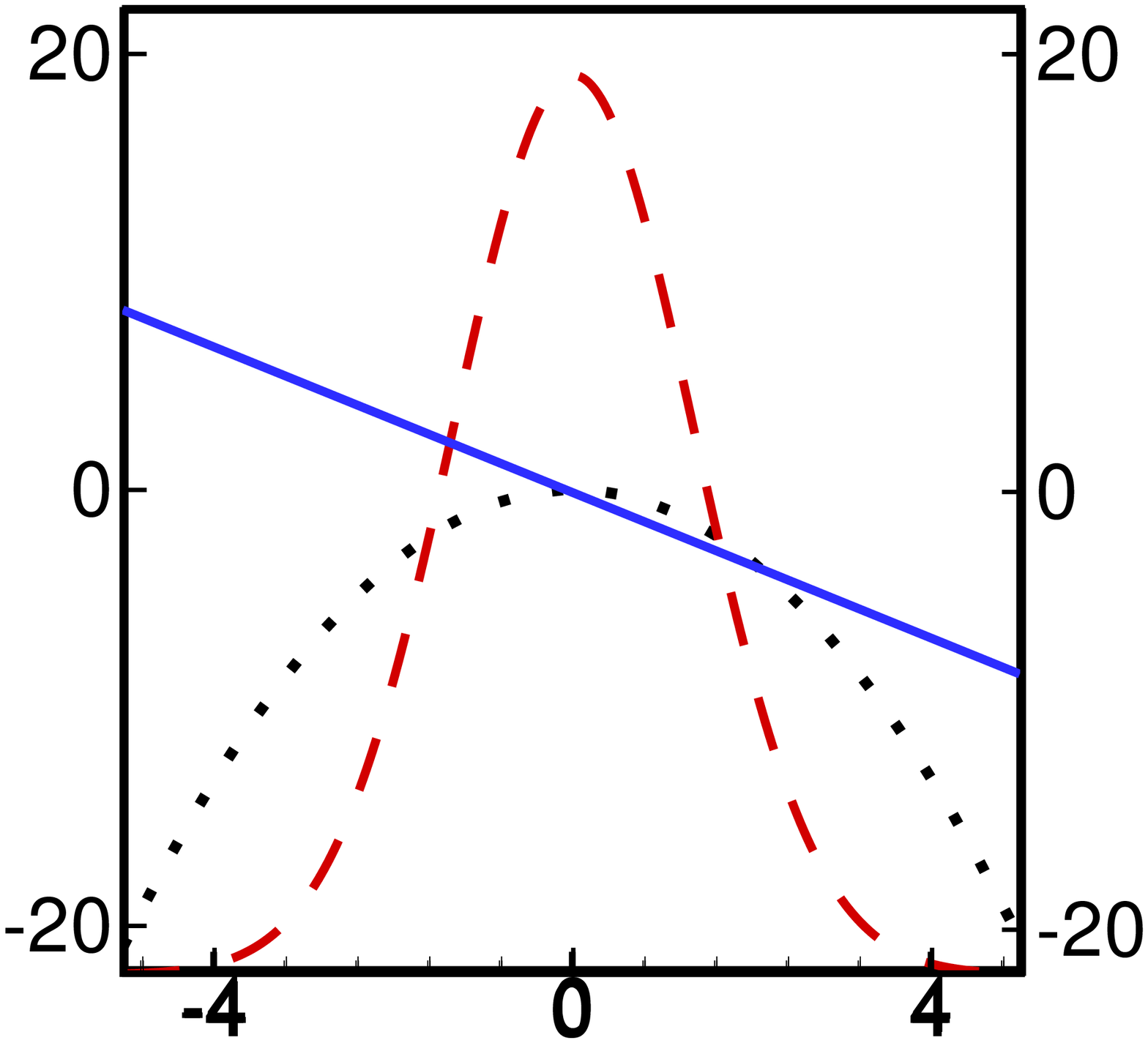}
   \includegraphics[height=0.42in]{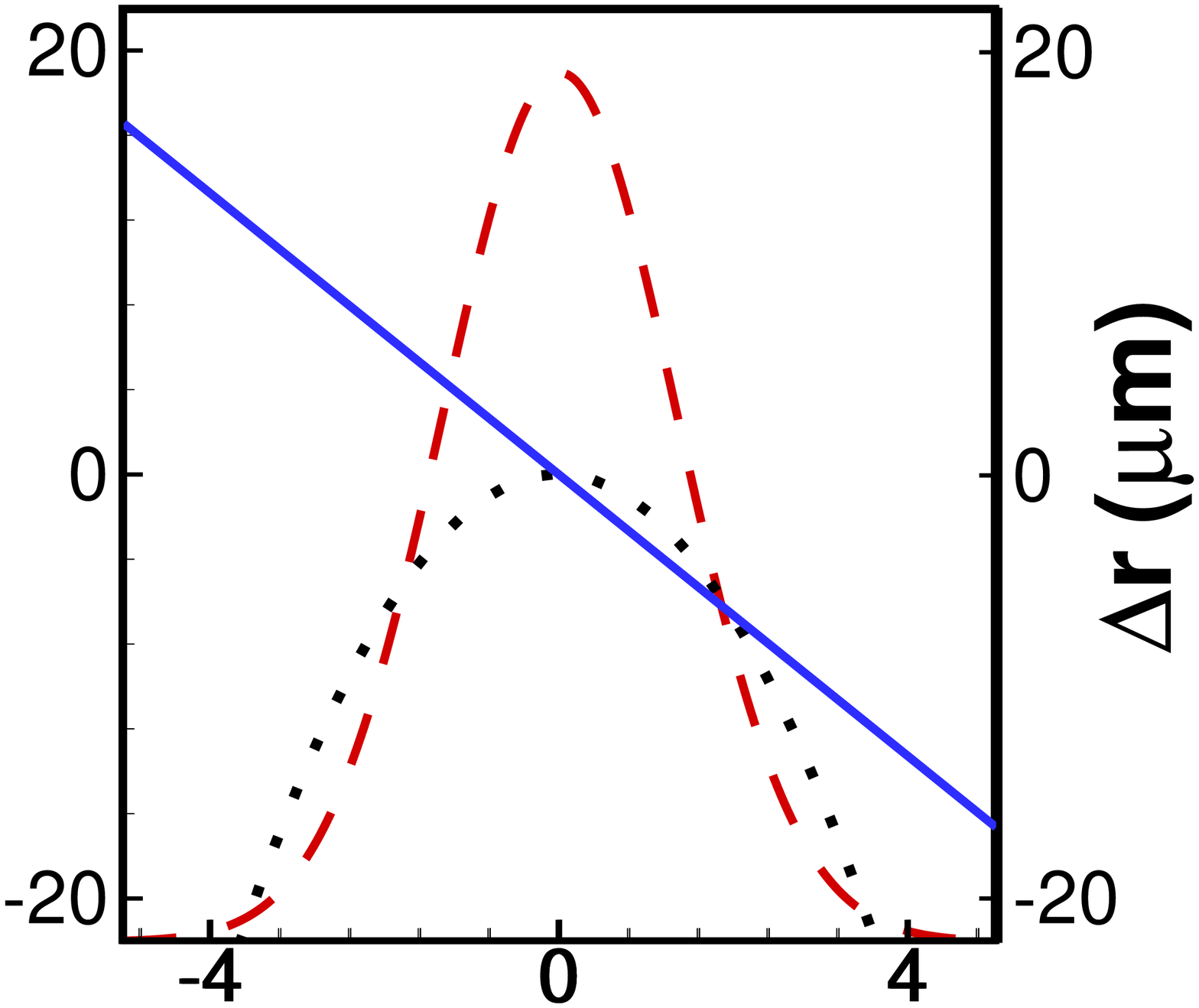}\\
   \includegraphics[height=0.45in]{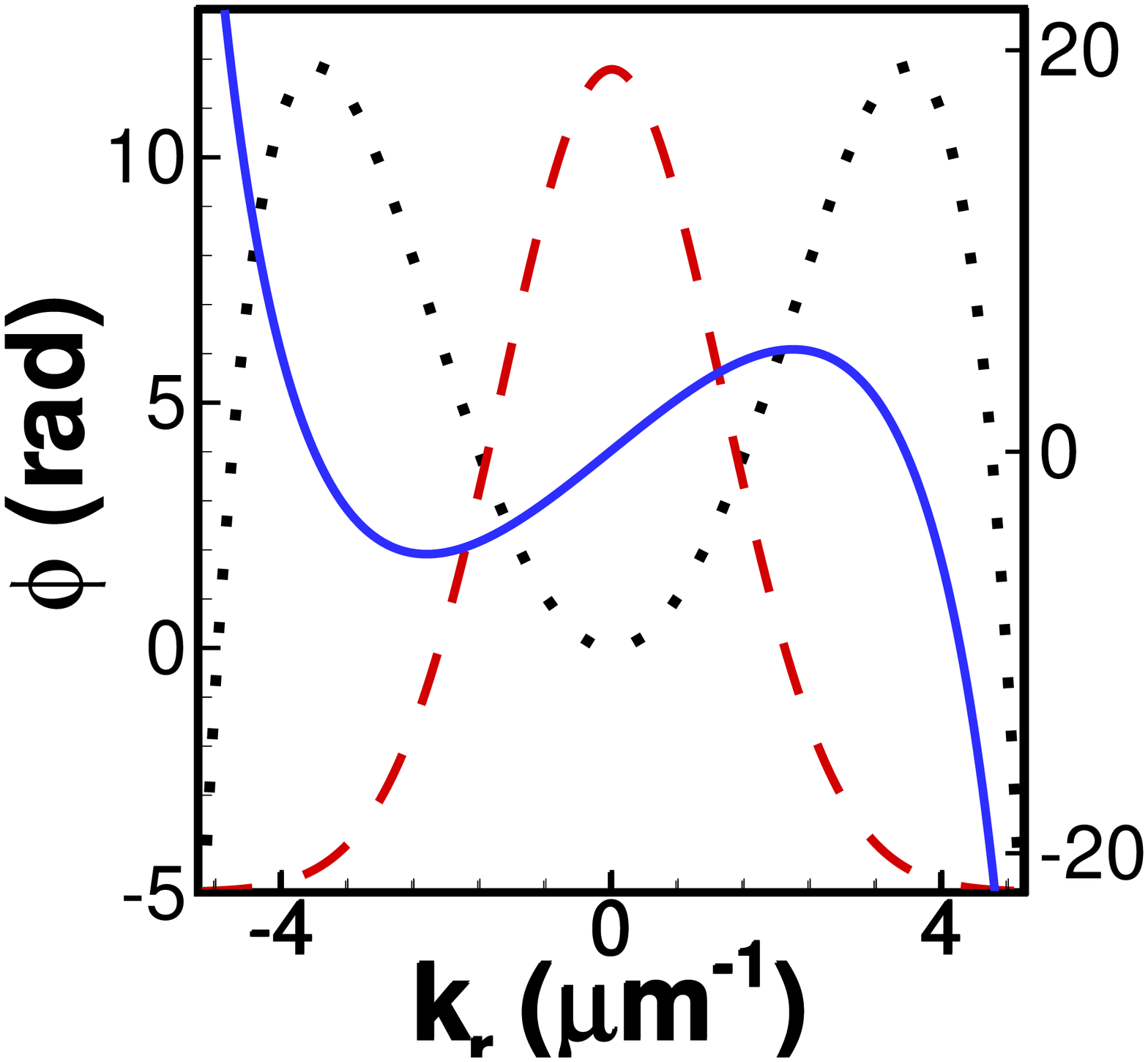}
   \includegraphics[height=0.45in]{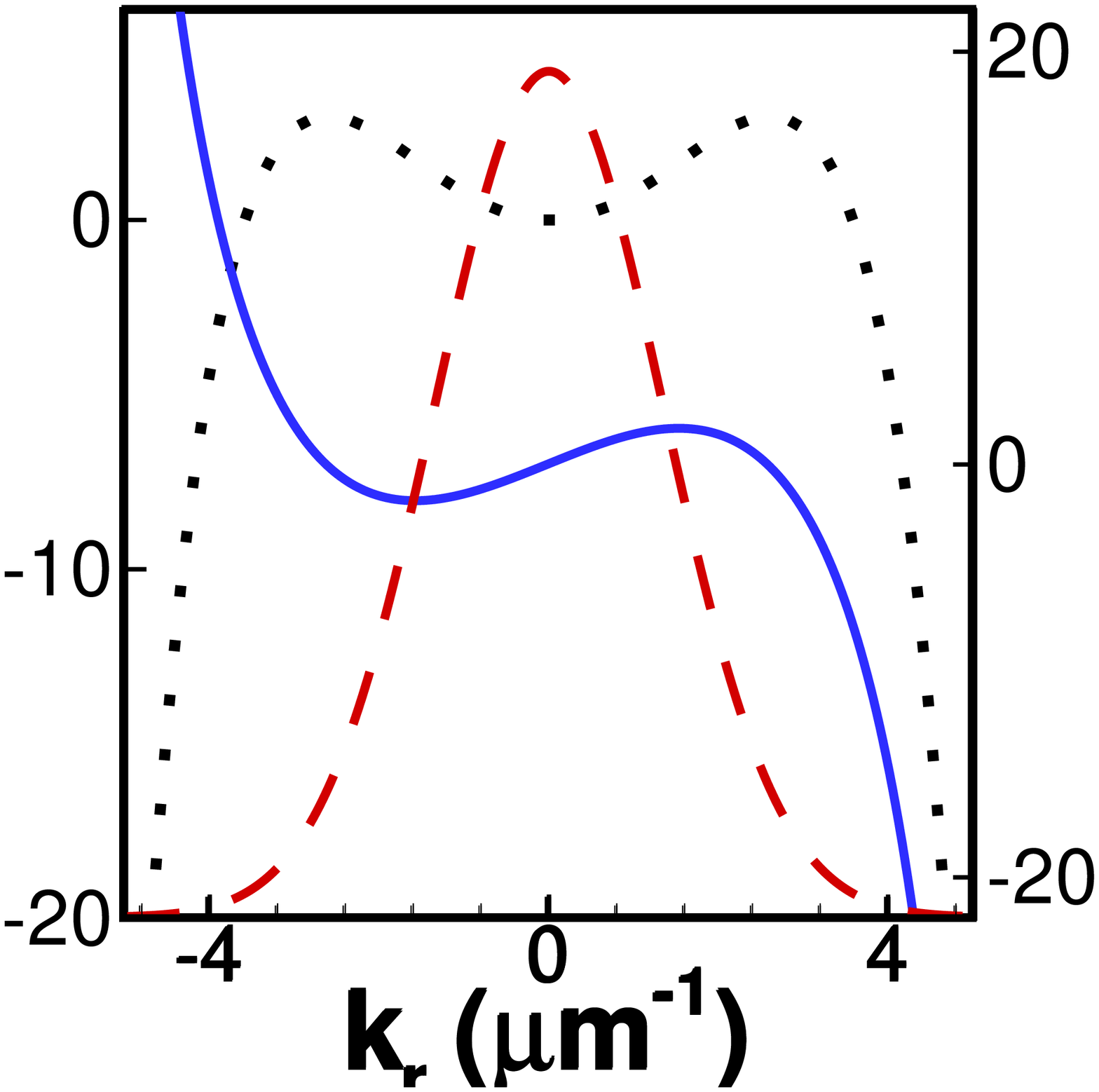}
   \includegraphics[height=0.45in]{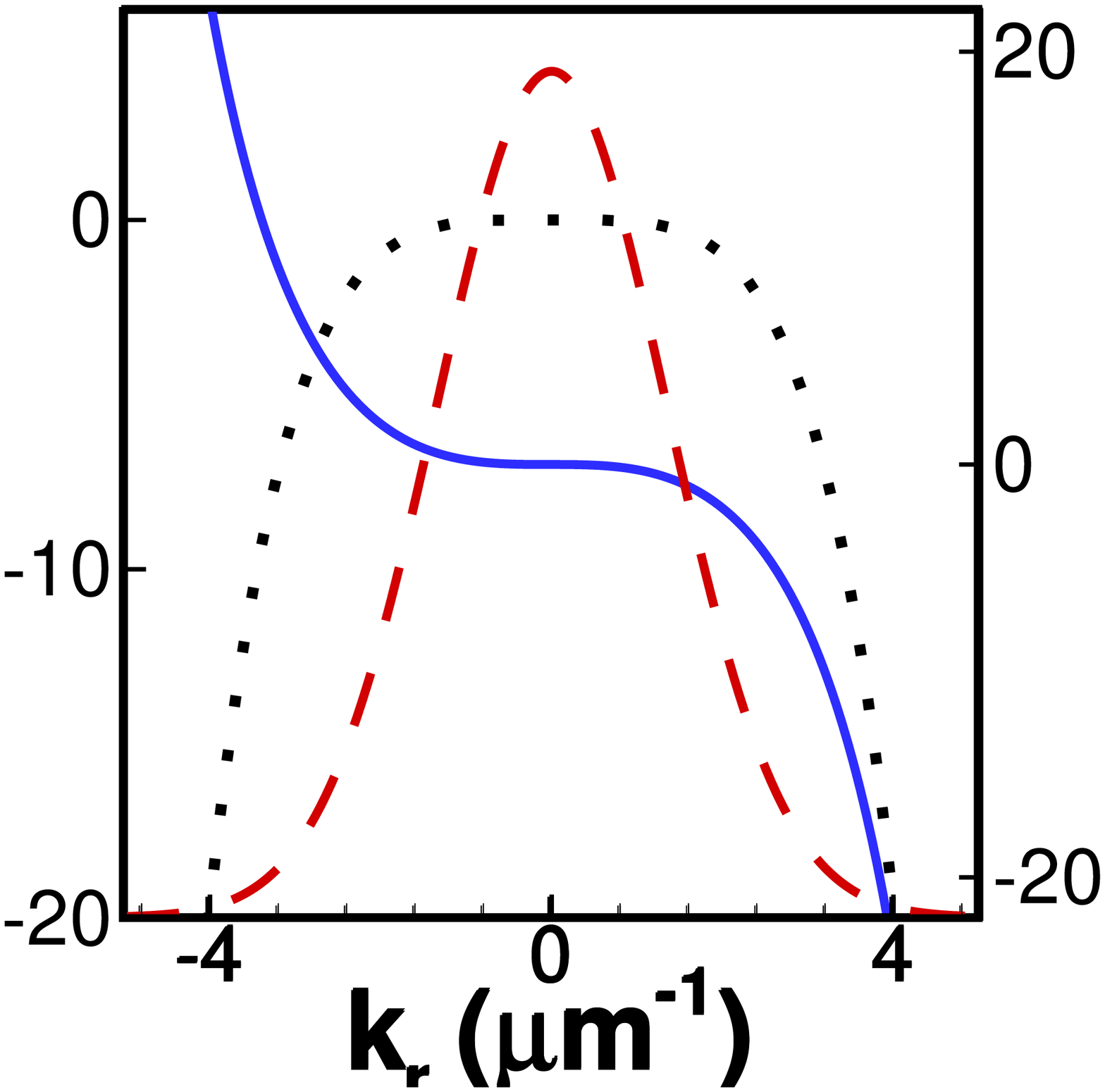}
   \includegraphics[height=0.45in]{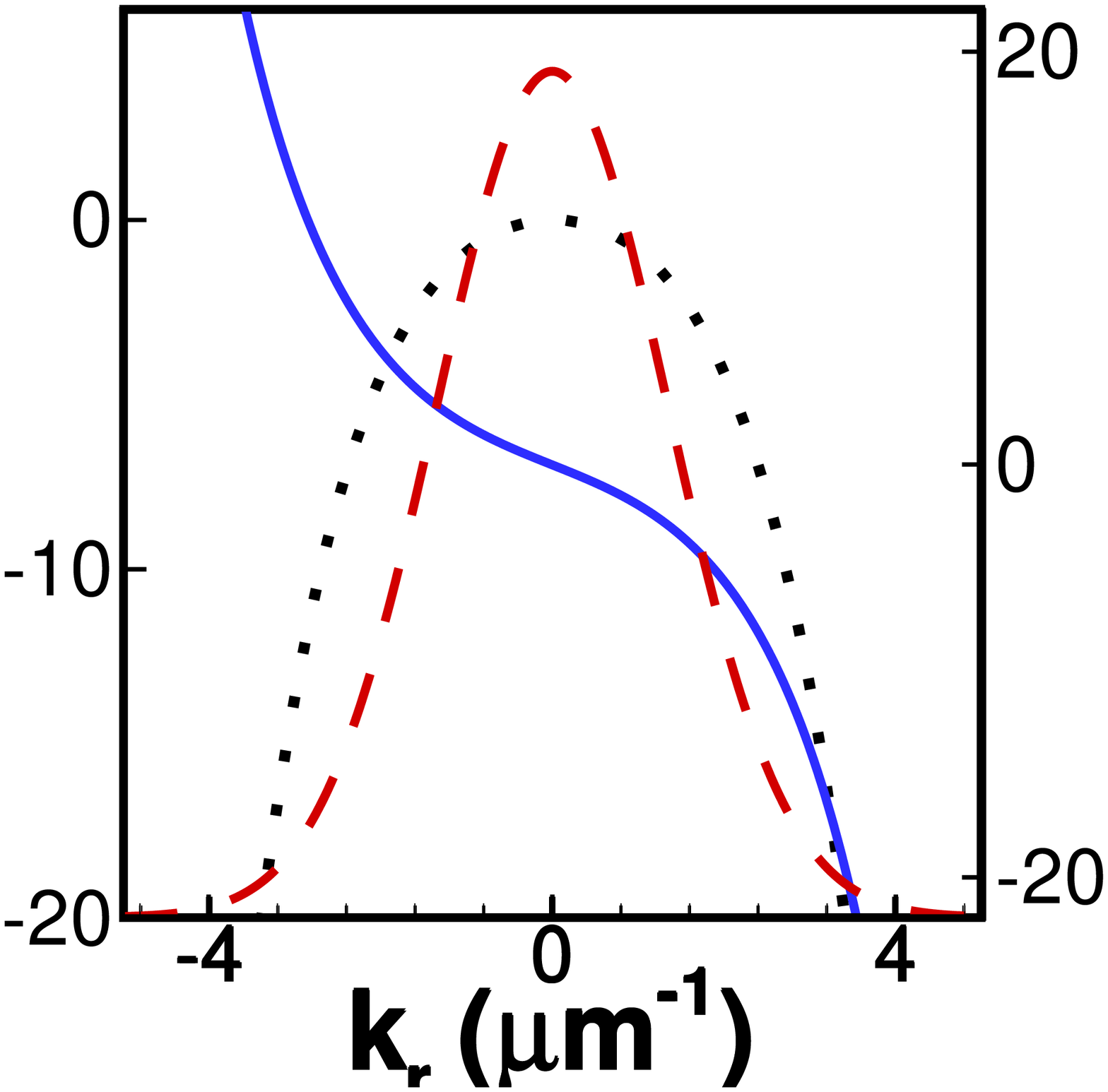}
   \includegraphics[height=0.45in]{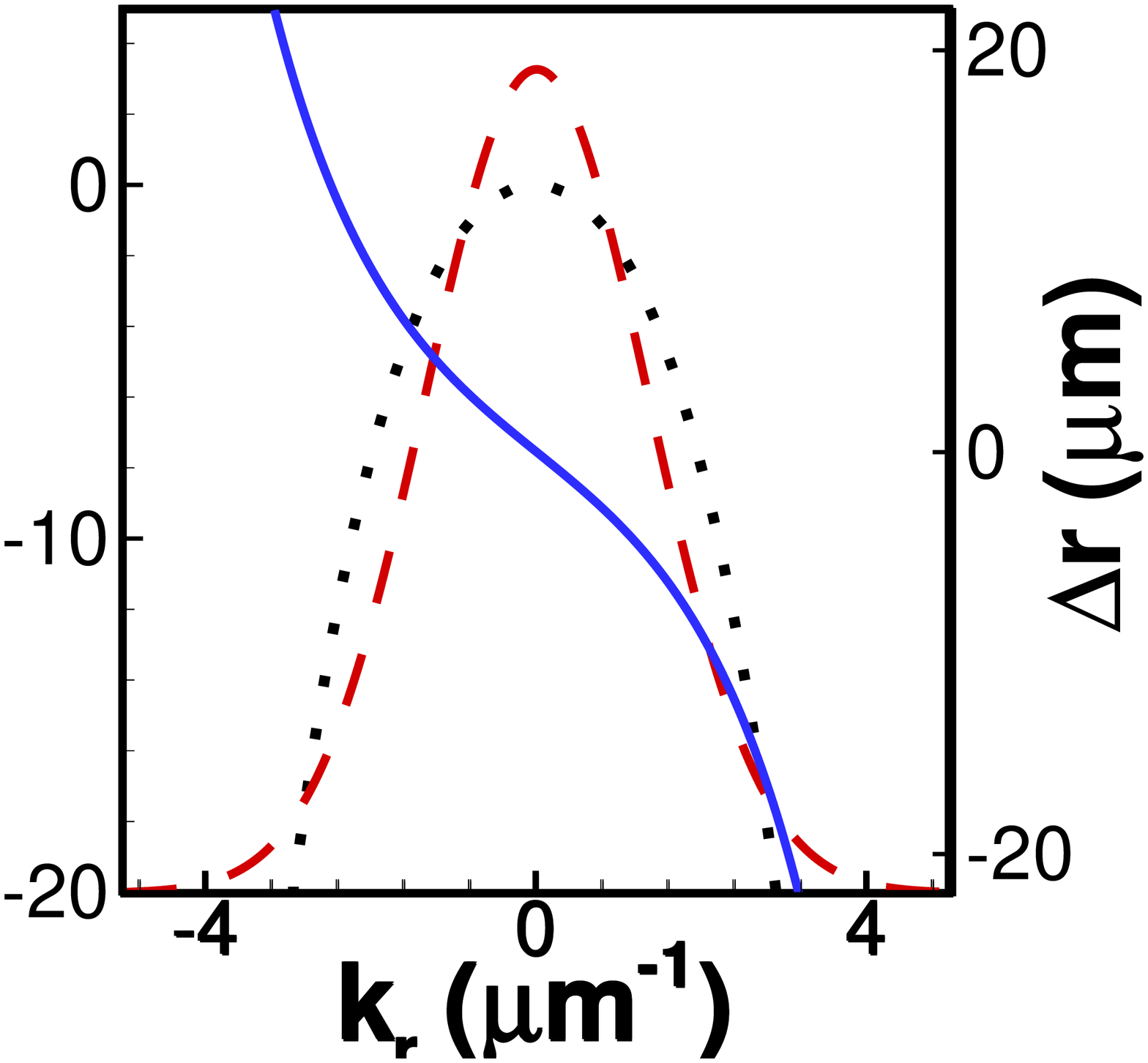}
  \end{center}
\caption{ Row 1: the evolution of the intensity distribution of a
tightly focused Gaussian beam in real domain resulted from the
paraxial theory. Row 2: the evolution of the intensity distribution
resulted from the nonparaxial theory. Row 3: The evolution of the
intensity and the phase  in the angular spectrum domain resulted
from the paraxial theory. Row 4: The evolution of the intensity and
the phase in the angular spectrum domain resulted from the
nonparaxial theory. The focusing plane is located at $z=-30z_R$, the
beam width at the waist plane resulted from the paraxial theory is
$1\mu m$, identical to the wavelenth.}
 \end{figure}

  \begin{figure}[t]
\begin{center}
   \leavevmode
   \includegraphics[height=2in]{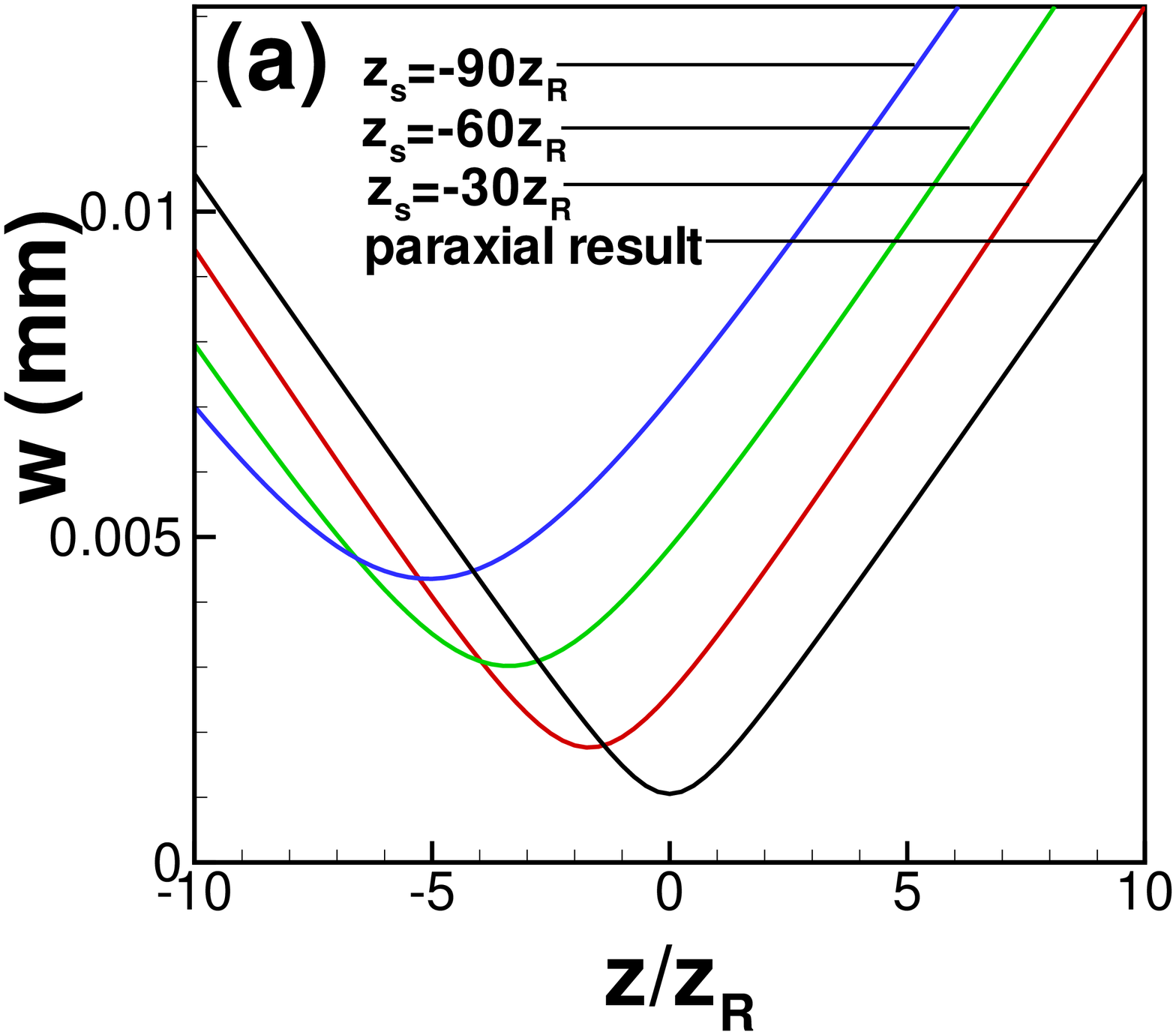}
   \includegraphics[height=2in]{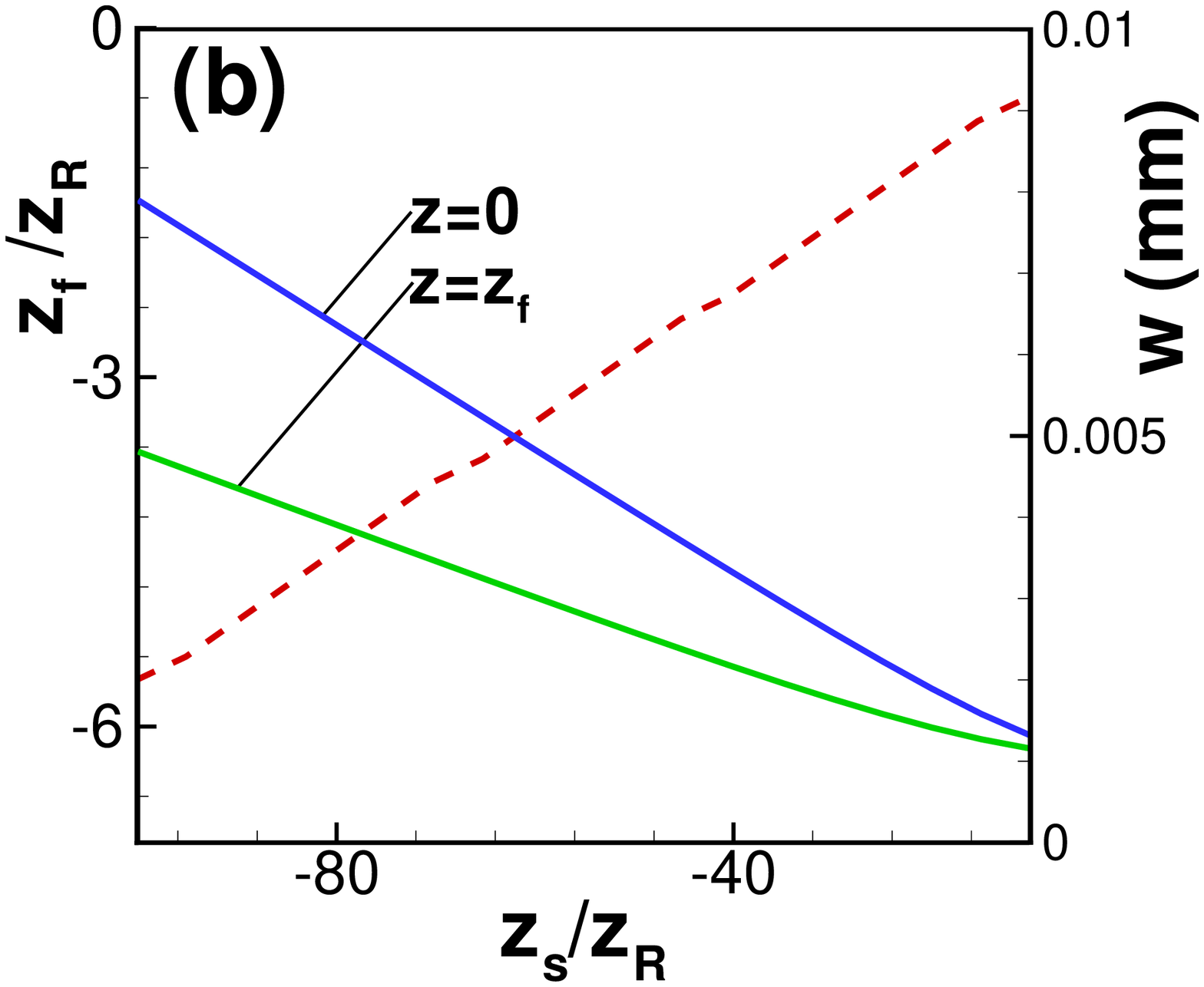}
  \end{center}
\caption{ (a) The comparison between the evolution of the beam width
resulted from the nonparaxial theory and from the paraxial theory
for various location of the focusing plane ($z_s$).
 (b) Dashed line: the distance between the real-waist ($z=0$)
and pseudo-waist ($z_f$) vs the  location of the focusing plane
($z_s$). Solid lines: the beam width at the real-waist ($z=0$)  and
pseudo-waist ($z=z_f$) vs the  location of the focusing plane
($z_s$). }
 \end{figure}

However, in fact the paraxial theory becomes invalid for the tightly
focused beams, because under this condition  the higher-orders of
spatial dispersion play critical roles in the evolution of the beam.
A tightly focused beam presents interesting evolution which is
critically different to the paraxial anticipation:
\\
 I) In the view of
Fourier optics, the higher-order spatial dispersion induces  the
addition of the fourth- and higher£­order phase factor in the
angular spectrum domain. mathematically, each Hermite-, Laguerre-,
or Ince-Gaussian functions with such a  phase factors are no longer
the eigen functions of Fourier transform. Therefore any paraxially
shape-invariant beams becomes shape-variant under the nonparaxial
condition (e.g. the tightly focused fundmental Gaussian beam Shown
in Fig. 1).
\\
II) At the pseudo-waist, the linear initial positive chirp only
balances the 2nd-order-spatial-dispersion-induced  negative chirp.
The higher-orders of spatial dispersion, which induces a nonlinear
negative chirp, would broadens the beam.The beam width is larger
than the Fourier-Transform-limited one predicted by the paraxial
theory. The farther the focusing plane is from the pseudo-waist, the
larger the chirp will be, and in turn the larger the beam width will
be resulted (Fig. 2).
\\
 III) There is  a plane where the chirp-induced broadening of beam width is
smallest (which is still larger than the Fourier-Transform-limited
one). If we call this real-waist, the real-waist becomes farther and
father from the pseudo-waist with the increase of the distance
between the focusing plane and the pseudo-waist (Fig. 2).
\\
IV) The chirp is asymmetric about the pseudo-waist, therefore
neither the pattern nor the beam width is symmetric about the
pseudo-waist. For the Gaussian beam, for example, at the planes
after the pseudo-waist the phase in angular spectrum domain
decreases monotonically with $|r|$ and result in an negative and
nonlinear chirp. Therefore the beam is wider than what predicted by
the paraxial theory. At some planes before the pseudo-waist, the
phase in angular spectrum domain dose not vary monotonically with
$|r|$ and results in an s-like distribution of the chirp. Therefore,
at such a plane there is the same chirp occurs at two values of
$k_\bot$. These two angular spectrum interfere constructively or
destructively, depending on their relative phase difference. The
interference then results in a multi-ringed distribution of the beam
in the spatial domain.

\begin{figure}[t]
\begin{center}
   \leavevmode
   \includegraphics[height=2in]{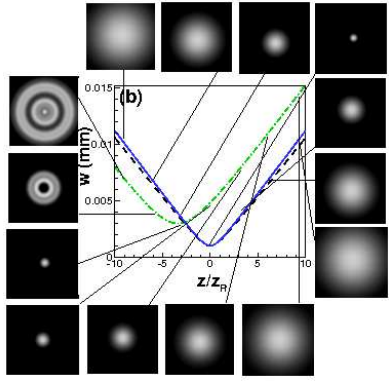}
    \end{center}
\caption{Dashed line and Dash-dotted line: the evolution of the beam
width of the focused Gaussian beam with no pre-added-chirp resulted
from the paraxial and nonparaxial theory, respectively. Solid line:
the evolution of the beam width of the chirp-pre-added focused
Gaussian beam resulted from the nonparaxial theory. The distance
between the focusing plane and the pseudo-waist plane is $\Delta
z=60z_R$.}
 \end{figure}

Although these phenomena are interesting, they frequently  occur
negative impact on real applications of the tightly focused beams.
In many applications, such as the laser-electron acceleration and
the trapping and manipulation of atom, the tightly focused beam is
expected to have the following properties: i)the spot size is the
smallest one to have the highest intensity; ii)the real-waist is
locate at the pseudo-waist, for the convenience of the controling of
the interaction. iii) the pattern at the real-waist is the same as
that at the focusing plane, for the convenient of the beam shaping.
Then a question arises: can we design the evolution of the tightly
focused beam to satisfy the above requirements? The answer is yes.
In fact, we can do this by modulate the forth- and higher-order
spatial dispersion. For example, if the field at the focusing plane
is the distance between the focusing plane and the pseudo-waist is
$\triangle z$, we can pre-add a chirp on the beam at the focusing
plane, i.e.
\begin{equation}
\tilde{E}^\prime(0)=\tilde{E}(0) \exp\left(-\frac{1}{4!}\beta_4
k_\bot^4\triangle z\right).
\end{equation}
then the field at the pseudo-waist plane becomes
\begin{eqnarray}
{E}^{(np)}(\triangle z)&=&
\int\int\tilde{E}^\prime(0)\exp\left[i\mathbf {k}_\bot \cdot \mathbf
{r}_\bot+\left(\beta_0+\frac{1}{2!}\beta_2
k_\bot^2+\frac{1}{4!}\beta_4 k_\bot^4\right) \Delta z\right]dk_xdk_y
\\
&=&\int\int\tilde{E}(0)\exp\left[i\mathbf {k}_\bot \cdot \mathbf
{r}_\bot+\left(\beta_0+\frac{1}{2!}\beta_2 k_\bot^2\right) \Delta
z\right]dk_xdk_y,
 \end{eqnarray}
 which is identical to
that resulted from the paraxial propagation. Therefore, if ${E}(0)$
is the  Hermite-, laguerre-, or Ince-Gaussian beams or the linear
superposition of them, the shape of the field at the pseudo-waist
plane is the same as ${E}(0)$. The width is the minimum; and the
pseudo-waist then becomes the real-waist (Fig. 3).

For a tightly focused beam, the beam width can be focused to the
size of $\mu m$ or even sub-$\mu m$. For such a narrow beam, even
the knife-edge based beam
 profilers is difficult to directly get the precise intensity distribution.
 On the other hand, If a traditional 4f system, which works well only in the paraxial
 condition, is directly introduced to amplify the
 pattern, the nonparaxiality would made the detected result deviate from the real
 intensity distribution. But based on the higher-order phase modulation, just like
 the method in the above paragraph, we can get the shape invariant and
 amplified intensity distribution. As a result, even a charge-coupled-device camera
can precisely detect the intensity distribution through this
approach.

In conclusion, the evolution of the tightly focused beams can be
explained with the theory of Fourier optics. In the propagation of
the tightly focused beam, the phase distribution in the angular
spectrum domain plays  an important role and induces interesting
beam patterns. By modulating the phase distribution in the angular
spectrum domain, one can made the field at the pseudo-waist plane
identical to that resulted from the paraxial propagation, of which
the width is the minimum and the pseudo-waist is located at the
real-waist. In the same approach, the intensity distribution of the
focused field can remain shape invariant and be
 amplified  so that even a charge-coupled-device camera
can precisely detect the intensity distribution.

\end{document}